\documentclass[12pt]{iopart}
\usepackage{graphicx}
\expandafter\let\csname equation*\endcsname\relax
\expandafter\let\csname endequation*\endcsname\relax

\usepackage{amsmath}
\usepackage{amssymb}

\begin{document}
\pagenumbering{roman}

\title{Transient force effects, as
predicted by Mbelek and Lachi\`{e}ze-Rey scalar tensor theory of gravitation}
\author{F. O. Minotti$^1$ and T. E. Raptis$^2$}
\address{$^1$Departamento de F\'{\i}sica, Facultad de Ciencias Exactas and Naturales,Universidad de Buenos Aires - Instituto de F\'{\i}sica del Plasma (CONICET), Buenos Aires, Argentina\\$^2$Division of Applied Technologies\\National Center for Science and Research "Demokritos", Athens, Greece.}

\ead{$^1$minotti@df.uba.ar, $^2$rtheo@dat.demokritos.gr}

\begin{abstract}
The scalar-tensor theory of gravitation proposed by Mbelek and Lachi\`{e}ze-Rey have been recently shown to lead to some remarkable effects beyond those initially explored by its authors. These new effects include a possible explanation of the forces measured in asymmetric resonant microwave cavities, and the variation of the amplitude of electromagnetic waves as they propagate in static electric or magnetic fields. These rather unique effects are excelent candidates for laboratory tests of this particular type of scalar-tensor theory. In the present work we introduce an additional remarkable effect of the theory: the generation of pulsed gravitational forces by transient, quasi-stationary electromagnetic fields. In particular, we explore the possible measurable effects of the simple experiment of turning on and off the current in a coil. We show that with the proposed values of the constant in the theory, this effect could resonantly excite a pendulum oscillation up to an easily measurable magnitude.          
\end{abstract}


\maketitle

\section{Introduction}

The scalar-tensor (ST) gravitational theory of Mbelek and Lachi\`{e}ze-Rey (MLR) \cite{MLR}, has the particularity of allowing electromagnetic (EM) fields to modify the space-time metric far more
strongly than predicted by General Relativity and standard ST theories. The theory was
applied in cosmological \cite{mbelek2003} and galactic \cite{mbelek2004}
contexts, and in \cite{MLR} it was used to explain the discordant measurements of Newton gravitational constant as due to the effect of the
Earth's magnetic field. In \cite{minotti} it was further shown that a ST theory
of the MLR type can explain the unusual forces on asymmetric resonant
cavities recently reported \cite{juan}. Besides, it was shown in \cite{raptis} that MRL theory also predicts rather strong effects on the amplitude of electromanetic waves propagation in static electric or magnetic fields. These unique effects area excelent candidates for table top experiments able to test this ST theory of gravitation.
We present in this work and additional effect predicted by MRL theory for quasi-stationary EM fields, the generation of transient gravitational effects of relatively large magnitude by transient EM fields. In particular, we explore with some detail the effect of turning or and off the current in a solenoid. Although we do not perform an exhaustive analytical analysis of this case, the expected transient source of the gravitational potential is shown to be able to resonantly excite the oscillation of a pendulum up to measurable amplitudes. 

\section{The MLR ST theory}

The details of the considered MLR ST theory where already presented in \cite{minotti} and \cite{raptis}, so that only a brief account will be given here. 
The MRL action is given by (SI units are used) 
\begin{eqnarray}
S &=&-\frac{c^{3}}{16\pi G_{0}}\int \sqrt{-g}\phi Rd\Omega +\frac{c^{3}}{%
16\pi G_{0}}\int \sqrt{-g}\frac{\omega \left( \phi \right) }{\phi }\nabla
^{\nu }\phi \nabla _{\nu }\phi d\Omega  \notag \\
&&+\frac{c^{3}}{16\pi G_{0}}\int \sqrt{-g}\phi \left[ \frac{1}{2}\nabla
^{\nu }\psi \nabla _{\nu }\psi -U\left( \psi \right) -J\psi \right] d\Omega 
\notag \\
&&-\frac{\varepsilon _{0}c}{4}\int \sqrt{-g}\lambda \left( \phi \right)
F_{\mu \nu }F^{\mu \nu }d\Omega -\frac{1}{c}\int \sqrt{-g}j^{\nu }A_{\nu
}d\Omega  \notag \\
&&+\frac{1}{c}\int \mathcal{L}_{mat}d\Omega .  \label{SKK}
\end{eqnarray}%
In (\ref{SKK}) the internal, non-dimensional scalar field is $\phi $, while $\psi $ is an external scalar field introduced by MRL to stabilize the reduced action of Kaluza-Klein multidimensional theories. These fields have vacuum expectation
values (VEV) $\phi _{0}=1$ and $\psi _{0}$, respectively. $G_{0}$ represents
Newton gravitational constant, $c$ is the velocity of light in vacuum, and $%
\varepsilon _{0}$ is the vacuum permittivity. $\mathcal{L}_{mat}$ is the
lagrangian density of matter. The other symbols are also conventional, $R$
is the Ricci scalar, and $g$ the determinant of the metric tensor $g_{\mu
\nu }$. The Brans Dicke parameter $\omega \left( \phi \right) $ is considered a
function of $\phi $. The function $\lambda \left( \phi \right) $ in the term of the action of the EM field is
found in reduced, effective theories \cite{mbelek2003}. The EM tensor is $F_{\mu \nu }=\nabla _{\mu }A_{\nu
}-\nabla _{\nu }A_{\mu }$, \ given in terms of the EM quadri-vector $A_{\nu
} $, with sources given by the quadri-current $j^{\nu }$. $U$ and $J$ are,
respectively, the potential and source of the field $\psi $. The source $J$
contains contributions from the matter, EM field and the scalar $\phi $. The
model for $J$ proposed in \cite{MLR} is 
\begin{equation}
J=\beta _{mat}\left( \psi ,\phi \right) \frac{8\pi G_{0}}{c^{4}}%
T^{mat}+\beta _{EM}\left( \psi ,\phi \right) \frac{4\pi G_{0}\varepsilon _{0}%
}{c^{2}}F_{\mu \nu }F^{\mu \nu },  \label{source}
\end{equation}%
where $T^{mat}$ is the trace of the energy-momentum tensor of matter,\ 
\begin{equation*}
T_{\mu \nu }^{mat}=-\frac{2}{\sqrt{-g}}\frac{\delta \mathcal{L}_{mat}}{%
\delta g^{\mu \nu }}.
\end{equation*}
The $\beta $ coefficients are unknown functions of the scalars, but in the weak-field (WF) approximation they
only contribute through the values of their first-order derivatives at the
VEV $\phi _{0}$ and $\psi _{0}$, and thus appear as adjustable constants.

Variation of (\ref{SKK}) with respect to $g^{\mu \nu }$ results in ($T_{\mu
\nu }^{EM}$ is the usual electromagnetic energy tensor)
\begin{eqnarray}
\phi \left( R_{\mu \nu }-\frac{1}{2}Rg_{\mu \nu }\right) &=&\frac{8\pi G_{0}%
}{c^{4}}\left[ \lambda \left( \phi \right) T_{\mu \nu }^{EM}+T_{\mu \nu
}^{mat}\right] +T_{\mu \nu }^{\phi }  \notag \\
&&+\frac{\phi }{2}\left( \nabla _{\mu }\psi \nabla _{\nu }\psi -\frac{1}{2}%
\nabla ^{\gamma }\psi \nabla _{\gamma }\psi g_{\mu \nu }\right)  \notag \\
&&+\frac{\phi }{2}\left( U+J\psi \right) g_{\mu \nu },  \label{Glm}
\end{eqnarray}
where $T_{\mu \nu }^{\phi }$ is the energy tensor of  the scalar $\phi$  
\begin{equation*}
T_{\mu \nu }^{\phi }=\nabla _{\mu }\nabla _{\nu }\phi -\nabla ^{\gamma
}\nabla _{\gamma }\phi g_{\mu \nu }+\frac{\omega \left( \phi \right) }{\phi }%
\left( \nabla _{\mu }\phi \nabla _{\nu }\phi -\frac{1}{2}\nabla ^{\gamma
}\phi \nabla _{\gamma }\phi g_{\mu \nu }\right) . 
\end{equation*}

Variation with respect to $\phi $ gives 
\begin{eqnarray*}
\phi R+2\omega \nabla ^{\nu }\nabla _{\nu }\phi &=&\left( \frac{\omega }{%
\phi }-\frac{d\omega }{d\phi }\right) \nabla ^{\nu }\phi \nabla _{\nu }\phi -%
\frac{4\pi G_{0}\varepsilon _{0}}{c^{2}}\phi \frac{d\lambda }{d\phi }F_{\mu
\nu }F^{\mu \nu } \\
&&-\frac{\partial J}{\partial \phi }\psi \phi +\phi \left[ \frac{1}{2}\nabla
^{\nu }\psi \nabla _{\nu }\psi -U\left( \psi \right) -J\psi \right] ,
\end{eqnarray*}
which can be rewritten, using the contraction of (\ref{Glm}) with $g^{\mu
\nu }$ to replace $R$, as 
\begin{eqnarray}
\left( 2\omega +3\right) \nabla ^{\nu }\nabla _{\nu }\phi &=&-\frac{d\omega 
}{d\phi }\nabla ^{\nu }\phi \nabla _{\nu }\phi -\frac{4\pi G_{0}\varepsilon
_{0}}{c^{2}}\phi \frac{d\lambda }{d\phi }F_{\mu \nu }F^{\mu \nu }+\frac{8\pi
G_{0}}{c^{4}}T^{mat}  \notag \\
&&+\phi \left[ \frac{1}{2}\nabla ^{\nu }\psi \nabla _{\nu }\psi -U\left(
\psi \right) -J\psi \right] -\frac{\partial J}{\partial \phi }\psi \phi ,
\label{phi}
\end{eqnarray}
where it was used that $T^{EM}=T_{\mu \nu }^{EM}g^{\mu \nu }=0$.

The non-homogeneous Maxwell equations are obtained by varying (\ref{SKK})
with respect to $A_{\nu }$, 
\begin{equation}
\nabla _{\mu }\left\{ \lambda \left( \phi \right) F^{\mu \nu }\right\} =\mu
_{0}j^{\nu }.  \label{Maxwell}
\end{equation}%
with $\mu _{0}$ the vacuum permeability.

Finally, the variation with respect to $\psi $ results in 
\begin{equation}
\nabla ^{\nu }\nabla _{\nu }\psi +\frac{1}{\phi }\nabla ^{\nu }\psi \nabla
_{\nu }\phi =-\frac{\partial U}{\partial \psi }-J-\frac{\partial J}{\partial
\psi }\psi +\frac{1}{\phi }\frac{8\pi G_{0}}{c^{4}}T^{mat}.  \label{psi}
\end{equation}

\section{Weak-field approximation}

As explained in \cite{mbelek2003} and \cite{mbelek2004}, the condition of
recovering Einstein-Maxwell equations when the scalar fields are not excited
requires that the $\beta $ coefficients in (\ref{source}) and the potential $%
U\left( \psi \right) $ be all zero when evaluated at the VEV $\phi _{0}$ and 
$\psi _{0}$, and also that $\lambda \left( \phi _{0}\right) =1$.

With these considerations, Eq. (\ref{SKK}) can be approximated in the WF limit keeping only the lowest significant order in the perturbations $h_{\mu \nu }$ of the metric $g_{\mu
\nu }$ about the Minkowski metric $\eta _{\mu \nu }$, with signature
(1,-1,-1,-1), and of the scalar fields about their VEV $\phi _{0}$ and $\psi
_{0}$, as

\begin{equation}
-\eta ^{\gamma \delta }\partial _{\gamma \delta }\overline{h}_{\mu \nu }=2\left( \partial _{\mu \nu }\phi
-\eta ^{\gamma \delta }\partial _{\gamma \delta }\phi \eta _{\mu \nu
}\right) ,  \label{Gik0}
\end{equation}
with the Lorentz gauge 
\begin{equation}
\partial _{\gamma }\overline{h}_{\nu }^{\gamma }=0,  \label{LG}
\end{equation}
where  
\begin{equation*}
\overline{h}_{\mu \nu }\equiv h_{\mu \nu }-\frac{1}{2}h\eta _{\mu \nu },
\end{equation*}
with $\overline{h}=\eta ^{\mu \nu }\overline{h}_{\mu \nu }$.
As can be seen in \cite{minotti}, for slow moving neutral masses, the WF limit of the geodesic equation corresponds to motion in flat Minkowski
space-time under the action of a specific force (per unit mass) given by (Latin indices correspond to the spatial coordinates) 
\begin{equation}
f_{i}=-\frac{c^{2}}{4}\frac{\partial }{\partial x_{i}}\left( \overline{h}
_{00}+\overline{h}_{kk}\right) +c\frac{\partial \overline{h}
_{0i}}{\partial t}.  \label{forcepermass}
\end{equation}

Using Eqs. (\ref{Gik0}) and (\ref{forcepermass}) the gravitational field is thus represented in the WF approximation by a gravitational potential $\chi $ in flat Minkowski
space-time, whose determining equation with only scalar field sources is 
\begin{equation}
\square \chi =\frac{\partial ^{2}\phi }{\partial t^{2}}-\frac{c^{2}}{2}
\square \phi ,  \label{dalemchi}
\end{equation}

With respect to the equation for the scalar fields in the WF approximation, as was argued in \cite{minotti}, in order for the MLR theory to be consistent with the lack of strong
gravitational effects due to the magnetic field of the Earth, the non-linear
terms in Eqs. (\ref{phi}) and (\ref{psi}) should be kept, even in the WF approximation. This is because the laplacian terms in these equations can be zero for this particular type of source, so that the equalities are satisfied by the higher order terms.

Specifically, for a static magnetic field outside its
sources one can write $\mathbf{B}=\mathbf{\nabla }\Psi $, with $\nabla
^{2}\Psi =0$, so that, from Eqs. (\ref{phi}) and (\ref{psi}) for the static
case, one has
\begin{subequations}
\label{phipsinl}
\begin{eqnarray}
\left( 2\omega+3\right) \nabla ^{2}\phi +\omega _{0}^{\prime }\mathbf{%
\nabla }\phi \cdot \mathbf{\nabla }\phi -\frac{1}{2}\mathbf{\nabla }\psi
\cdot \mathbf{\nabla }\psi  &=&\chi _{\phi }\mathbf{\nabla }\Psi \cdot 
\mathbf{\nabla }\Psi , \\
\nabla ^{2}\psi +\mathbf{\nabla }\phi \cdot \mathbf{\nabla }\psi  &=&\chi
_{\psi }\mathbf{\nabla }\Psi \cdot \mathbf{\nabla }\Psi ,
\end{eqnarray}%
where $\omega _{0}^{\prime }\equiv \left( d\omega /d\phi \right) _{\phi _{0}}
$, and 
\end{subequations}
\begin{eqnarray*}
\chi _{\phi } &\equiv &\frac{8\pi G_{0}\varepsilon _{0}}{c^{2}}\psi
_{0}\left. \frac{\partial \beta _{EM}}{\partial \phi }\right\vert _{\phi
_{0},\psi _{0}}, \\
\chi _{\psi } &\equiv &\frac{8\pi G_{0}\varepsilon _{0}}{c^{2}}\psi
_{0}\left. \frac{\partial \beta _{EM}}{\partial \psi }\right\vert _{\phi
_{0},\psi _{0}}.
\end{eqnarray*}
These equations have the solutions $\mathbf{\nabla }\phi \propto \mathbf{%
\nabla }\psi \propto \mathbf{\nabla }\Psi $, so that $\nabla ^{2}\phi
=\nabla ^{2}\psi =0$, thus nullifying the contribution of the vacuum
magnetic field as a source of the gravitational potential in the static version of Eq. (\ref{dalemchi}). We note that the laplacian operator in  Eqs. (\ref{phipsinl}) is to be interpreted as 
\begin{equation*}
\nabla ^{2}\phi = -g ^{\mu \nu} \partial_{\mu \nu} \phi ,  
\end{equation*}
for the static case, and so Eqs. (\ref{phipsinl}) are correct up to order two in the perturbed fields.
This kind of solution for the case of the Earth's
magnetic field is compatible with the proposal in \cite{MLR}, where the
solution with $\nabla ^{2}\phi \neq 0$ was used, on the condition that 
\begin{equation}
\omega _{0}^{\prime }\sim 2\omega _{0}+3.  \label{condmlr}
\end{equation}

In this way, one has
\begin{equation*}
\mathbf{\nabla }\phi
\equiv \Gamma \mathbf{\nabla }\Psi,
\end{equation*}
where the constant $\Gamma $ depends, other than on $\lambda _{0}^{\prime }$%
, also on $\chi _{\phi }$, $\chi _{\psi }$ and $\omega _{0}^{\prime }$, and is
directly determined from the system (\ref{phipsinl}) with the condition $%
\nabla ^{2}\phi =\nabla ^{2}\psi =0$. From the data in \cite{MLR}, together
with the \textit{assumption} $\chi _{\psi }\approx \chi _{\phi }$, and
condition (\ref{condmlr}), one has 
\begin{equation}
\Gamma \simeq \sqrt{-\frac{8\pi G_{0}\varepsilon _{0}}{%
\left( 2\omega _{0}+3\right) c^{2}}\psi _{0}\left. \frac{\partial \beta _{EM}%
}{\partial \phi }\right\vert _{\phi _{0},\psi _{0}}}\approx  10^{-4}\frac{A}{N}.  \label{gammon}
\end{equation}

For the case of quasi-stationary fields, for which 
\begin{equation}
\frac{1}{c^{2}}\left( \frac{\partial \phi }{\partial t}\right) ^{2}\ll 
\mathbf{\nabla }\phi \cdot \mathbf{\nabla }\phi,   \label{quasi}
\end{equation}
with equal condition for the scalar $\psi$, Eqs. (\ref{phipsinl}) are correct up to order one in a small parameter representing the ratio of time derivatives to spatial derivatives ($a$ is a characteristic length of the system, and $\tau$ a characteristic time of variation)
\begin{equation*}
\epsilon = \frac{a}{c \tau}.
\end{equation*}
This is so because, as with the time derivatives in their left-hand sides, also the corrections in the right-hand sides of Eqs. (\ref{phipsinl}) are of order $\epsilon^{2}$.   

If one were simply to use this solution with $\nabla ^{2}\phi =0$ in (\ref{dalemchi}) the resulting equation would be
\begin{equation}
\nabla ^{2}\chi =-\frac{\Gamma }{2}\frac{\partial ^{2}\Psi }{\partial t^{2}}.
\label{chi_final}
\end{equation}
However, the retained term is of order $\epsilon^{2}$ compared with the neglected one, so that in principle, by including the missing order $\epsilon^{2}$ terms in Eqs. (\ref{phipsinl}), a non-zero correction of that order could be found for $\nabla ^{2}\phi$. Even for the simplest EM sources this task is extremely difficult to perform analytically. Besides, the existing uncertainties in the parameters of the theory do not allow a precise comparison of theory and experiment, but only order of magnitude estimations. In this way, at this point we will consider only the effect of the retained term in (\ref{chi_final}) as an indicator of the order of magnitude reasonably expected of the transient effect. 

\section{Transient magnetic field}

If a uniform magnetic field of slowly changing magnitude $B\left( t\right) $
exists inside a sphere of radius $a$, the magnetic field is given as (in
spherical coordinates)
\begin{equation*}
\mathbf{B}=\mathbf{\nabla }\Psi =\left\{ 
\begin{array}{c}
B\left( t\right) \mathbf{e}_{z}\hspace{1cm}r<a \\ 
\frac{a^{3}B\left( t\right) }{2r^{3}}\left( 2\cos \theta \mathbf{e}_{r}+\sin
\theta \mathbf{e}_{\theta }\right) \hspace{1cm}r>a%
\end{array}
\right. ,
\end{equation*}
so that%
\begin{equation*}
\Psi =\left\{ 
\begin{array}{c}
B\left( t\right) \,r\cos \theta \hspace{1cm}r<a \\ 
-\frac{a^{3}B\left( t\right) }{2r^{2}}\cos \theta \hspace{1cm}r>a
\end{array}%
\right. ,
\end{equation*}%
and, from (\ref{chi_final}), one has at leading order%
\begin{equation*}
\nabla ^{2}\chi \simeq \left\{ 
\begin{array}{c}
\frac{\Gamma B^{\prime \prime }\left( t\right) \,r}{2}\cos \theta \hspace{1cm}r<a \\ 
-\frac{\Gamma a^{3}B^{\prime \prime }\left( t\right) }{4r^{2}}\cos \theta 
\hspace{1cm}r>a%
\end{array}%
\right. ,
\end{equation*}%
where the primes indicate the time derivative. The solution of this equation
that is continuous up to first order derivatives in $r=a$ is%
\begin{equation*}
\chi =\left\{ 
\begin{array}{c}
\frac{\Gamma B^{\prime \prime }\left( t\right) }{20}\,r^{3}\cos \theta 
\hspace{1cm}r<a \\ 
\frac{\Gamma a^{3}B^{\prime \prime }\left( t\right) }{8}\left( 1-\frac{3a^{2}%
}{5r^{2}}\right) \cos \theta \hspace{1cm}r>a%
\end{array}%
\right. .
\end{equation*}

In this way, the rapidly dominating term for $r>a$ corresponds to a specific
force (per unit mass)%
\begin{equation}
\mathbf{f}=-\mathbf{\nabla }\chi \simeq \frac{\Gamma a^{3}B^{\prime \prime
}\left( t\right) }{8r}\sin \theta \mathbf{e}_{\theta }.  \label{force}
\end{equation}

\begin{figure}[ht]
\includegraphics[width=0.8\textwidth]{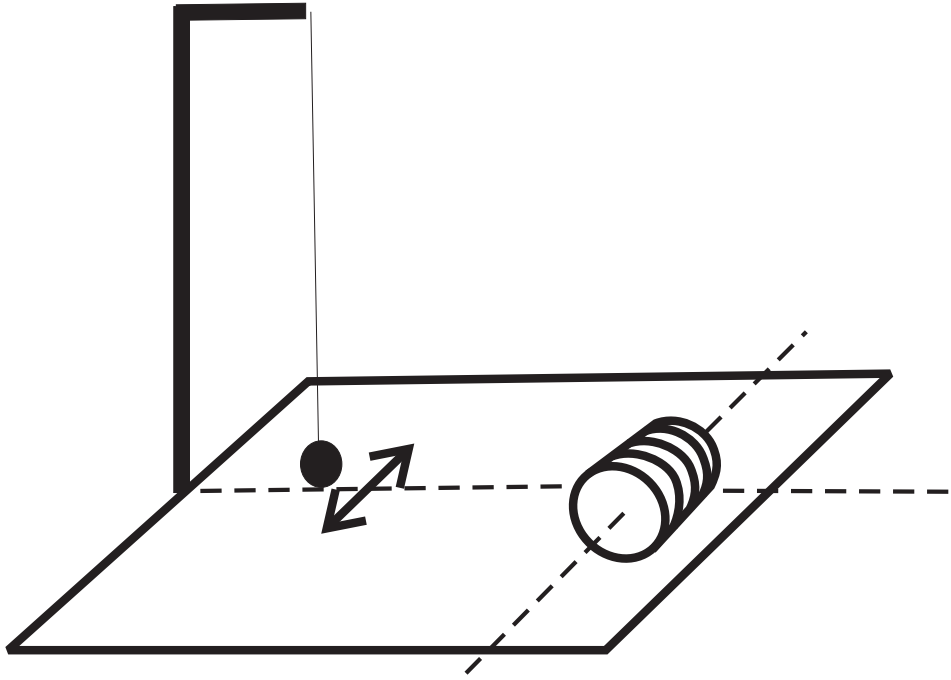}
\caption{Conceptual sketch showing the disposition of the diverse elements and the expected direction of movement of the pendulum.}
\label{fig1}
\end{figure}

The solution (\ref{force}) is valid also for a solenoid of linear dimension $%
a$ for $r\gg a$, so that we can very easily estimate the specific force for
the case of a solenoid whose current is switched either on or off. For a
linear, cylindrical solenoid of length $l$ and radius $a$, with $N$ turns
and magnetic permeability $\mu $ one has 
\begin{equation*}
B\left( t\right) =\frac{\mu N}{l}i\left( t\right) ,
\end{equation*}%
with self-inductance%
\begin{equation*}
L=\frac{\mu N^{2}\pi a^{2}}{l},
\end{equation*}%
and where the electric current $i\left( t\right) $ satisfies%
\begin{equation*}
i\left( t\right) =\frac{V}{R}\left( 1-\exp \left( -t/\tau \right) \right)
\end{equation*}%
when it is switched on, and 
\begin{equation*}
i\left( t\right) =\frac{V}{R}\exp \left( -t/\tau \right) ,
\end{equation*}%
when it is switched off, with $V$ the voltage of the power supply, $R$ the
circuit resistance and $\tau =L/R$. During the whole transient of the
turning on or off of the magnetic field, the force (\ref{force}) generates
an "impulsive" velocity on a moving mass of value%
\begin{equation*}
\Delta \mathbf{v}=\int\limits_{0}^{\infty }\mathbf{f\,}dt=-\frac{\Gamma
a^{3}B^{\prime }\left( t=0\right) }{8r}\sin \theta \mathbf{e}_{\theta }.
\end{equation*}%
Using the previous expressions it results that%
\begin{equation*}
\Delta \mathbf{v}=\mp \frac{\Gamma aV}{8\pi Nr}\sin \theta \mathbf{e}%
_{\theta },
\end{equation*}%
where the $-$ ($+$) sign corresponds to the switching on (off) of the
current in the solenoid.

If the moving mass is the bob of a pendulum, by switching on and off the
current to match the natural period of oscilation of the pendulum one could
obtain a resonant forcing with noticeable effects.
A sketch of the disposition of pendulum and coil for this kind of experiment can be seen in Fig. 1.

A simple simulation was performed of the motion of a pendulum with a linear friction model, and a periodic impulsive forcing (alternating in sign) with amplitude
\begin{equation*}
\Delta \mathbf{v}_{0}=\frac{\Gamma a V}{8\pi N r},
\end{equation*}
and period $\Omega$.
For low enough friction (about $100$ free oscillations for a $1/e$ decay of the amplitude) and $\Omega$ matching the natural frequency of the pendulum with relative error less than $5\%$, using a coil of $N = 5$ turns, $a/r = 0.1$, and length of the pendulum about $1 m$, sources values of a few $kV$ resulted in permanent amplitude oscillations of the order of $mm$, which should be easily appreciated.  

\section{Conclusions}

We have derived the equations for the expected gravitational effects of a transient quasi-stationary magnetic field, as predicted by MRL theory, and simulated the expected results in a simple enough experiment, which could easily discriminate the transient effects predicted.

\end{document}